\documentclass{article}
\usepackage{spconf,amsmath,graphicx}
\usepackage{amstext,amsthm,amssymb,curves,xcolor,cancel}
\usepackage[retainorgcmds]{IEEEtrantools}
\usepackage{epstopdf}

\def\R{\mathbb{R}}

\def\d{\displaystyle}

\theoremstyle{plain}
\newtheorem{thm}{Theorem}[section]
\theoremstyle{definition}

\theoremstyle{definition}
\newtheorem{defn}[thm]{Definition}
\theoremstyle{definition}

\theoremstyle{plain}

\theoremstyle{plain}

\theoremstyle{plain}
\newtheorem{lem}[thm]{Lemma}
\theoremstyle{plain}

\title{Performance Analysis of Joint-Sparse Recovery from Multiple Measurement Vectors with Prior Information via Convex Optimization}

\name{Shih-Wei Hu, Gang-Xuan Lin, Sung-Hsien Hsieh, Wei-Jie Liang, and Chun-Shien Lu}
\address{Institute of Information Science, Academia Sinica, Taipei, Taiwan}

\begin{document}

\maketitle

\begin{abstract}

We address the problem of compressed sensing with multiple measurement vectors associated with prior information in order to better reconstruct an original sparse signal.
This problem is modeled via convex optimization with $\ell_{2,1}-\ell_{2,1}$ minimization.
We establish bounds on the number of measurements required for successful recovery.
Our bounds and geometrical interpretations reveal that if the prior information can decrease the statistical dimension and make it lower than that under the case without prior information, $\ell_{2,1}-\ell_{2,1}$ minimization improves the recovery performance dramatically.
All our findings are further verified via simulations.

\end{abstract}

\section{Introduction}\label{sec:intro}
\subsection{Background and Problem Definition}
Compressive sensing (CS) \cite{Baraniuk07, CRT06, Donoho06} of sparse signals in achieving simultaneous data acquisition and compression has been extensively studied in the past few years.
In this paper, we focus on multiple measurement vectors (MMVs) that are sensing results with respect to observed signals.
MMVs gradually exhibit the applicability especially in the areas of wireless sensor networks and wearable sensors \cite{LTW15,Z15,XLLD14}.

Let $S=[s_1,s_2,...,s_l]\in\R^{n\times l}$ be the matrix of $l$ ($> 1$) original signals to be sensed by a sensing matrix $\Phi\in\R^{m\times n}(m<n)$ and let the matrix of measurement vectors be $Y=[y_1,y_2,...,y_l]\in\R^{m\times l}$, where $y_i=\Phi s_i$, $i=1,2,...,l$.
Suppose there exists a orthonormal basis $\Psi$ such that $s_i=\Psi x_i$ and $X_0=[x_1,x_2,...,x_l]\in\R^{n\times l}$ is $k$-joint sparse. In other words, all $x_i$'s share the common support.
Given $A=\Phi\Psi$, recovery from MMVs can be efficiently solved via convex optimization as:
\begin{equation*}
(\mbox{Mconvex})~~\d\min_X f(X)~\textrm{ s.t. }~Y=AX,
\end{equation*}
where $f(\cdot)$ denotes a convex function.
We call the problem (Mconvex) succeeds if it has a unique optimal solution and is ground truth $X_0$.
In this paper, the convex function is chosen as $f(X)=\|X\|_{2,1}$ to enhance the joint-sparsity of $X$:
\begin{equation*}
(\mbox{ML1})~~\displaystyle\min_X\|X\|_{2,1}~\textrm{ s.t. }~Y=AX.
\end{equation*}

So far, there is very limited literature about MMVs with prior information via convex optimization.
In fact, we can have some prior knowledge about the ground truth $X_0$ in, for example, the problem of distributed compressive video sensing (DCVS) \cite{KangLu09}.
In DCVS, we usually adopt higher/lower measurement rates to sample and transmit key/non-key frames at encoder, and then we treat these reconstructed key frames as the prior information for better recovery of the non-key frames at decoder.
Mota {\em et al.} \cite{MDR14} first propose the analysis of single measurement vector (SMV) with prior information via convex optimization.
They show that the performance can be improved provided good prior information can be available.
In \cite{HLHL15}, we characterize when problem (ML1) succeeds and derive the phase transition of success rate inspired by the framework of conic geometry \cite{ALMT14}.

In this paper, we further extend the problem (ML1) to (ML1) plus prior information as:
\begin{equation*}
(\mbox{ML1P})~~\displaystyle\min_X\|X\|_{2,1}+\lambda\|X-W\|_{2,1}~\textrm{ s.t. }~Y=AX,
\end{equation*}
where $W$ is prior information associated with ground truth $X_0$.
The goal here is to provide theoretical but practical bound of the probability of successful recovery
and analyze the relationship between prior information and performance.

\subsection{Contributions of This Paper}
We summarize the contributions of our works here.
\begin{enumerate}
\item[$\bullet$] Based on conic geometry, the phase transition of success rate in (ML1P) is derived and is consistent with the empirical results.
This study indeed provides the useful insights into how to solve the problem of MMVs with prior information.
\item[$\bullet$] What prior information is ``good'' can be concluded by our theoretical analysis.
For example, instead of giving the rough conclusion such as $\|X_0 - W\|_{2,1}$ being close to 0, we clearly show how the supports of $X-W$ and the signs of $X-W$ affect the performance.
\end{enumerate}

\subsection{Notations}\label{subsec:notation}
For a matrix $H$, we denote its transpose by $H^T$; its $i^{\scriptsize\mbox{th}}$ row by $h^i$; its $j^{\scriptsize\mbox{th}}$ column by $h_j$; and the $i^{\scriptsize\mbox{th}}$ entry of $j^{\scriptsize\mbox{th}}$ column by $h^i_j$.
$\Lambda_H:=\{i:\|h^i\|_2\neq 0\}$ for a matrix $H$ is a support set that collects the indices of nonzero rows of $H$.
$\left\|\cdot\right\|_p$ and $\left\|\cdot\right\|_F$ denote the $\ell_p$-norm and Frobenius norm, respectively.
The $\ell_{p,q}$-norm of a matrix is defined as $\|X\|_{p,q}=\|(\|x^i\|_p)_{n\times 1}\|_q.$
The null space of matrix $A\in\mathbb{R}^{m\times n}$ is defined as
$\mbox{null}(A,l)=\left\{Z\in\mathbb{R}^{n\times l}:AZ=\textbf{0}_{m\times l}\right\}$.
Let $\mathbb{E}$ denote the expected value and let $\overline{\mathcal{B}}=\left\{x:\left\|x\right\|_2\leq1,~x\in\R^n\right\}$ denote closed unit ball.
The dot product of two matrices is $\langle X,Y\rangle=\mbox{tr}\left(X^TY\right)$.

\section{Conic Geometry}\label{sec:conic_geometry}
We briefly introduce how a convex function can be specified in terms
of conic geometry to make this paper self-contained.
First, we introduce a cone and measure its size in a sense of statistical dimension.
Then,  they are connected with optimality condition for the MMVs recovery problem.
\begin{defn}
(Descent cone \cite{ALMT14})\\
The descent cone $\mathcal{D}(f,x)$ of a function $f:\R^n\rightarrow\overline{\R}$ at a point $x\in\R^n$, defined as:
$$\mathcal{D}(f,x):=\bigcup_{\tau>0}\{u\in\R^n:f(x+\tau u)\leq f(x)\},$$
is the conical hull of the perturbations that do not increase $f$ near $x$.
\end{defn}

By the definition of descent cone, the necessary and sufficient condition of the success of problem (ML1) is described and proved in our earlier work \cite{HLHL15}.
But in this paper, the main problem we are studying is not related to a norm function, so we need to modify the proof slightly to fit the problem (Mconvex) with general convex function as follows.

\begin{lem}\label{ground truth}
\emph{(Optimality condition for MMVs recovery with general convex function)}\\
The matrix $X_0$ is the unique optimal solution to problem (Mconvex) if and only if $\mathcal{D}(f,X_0)\cap\emph{\mbox{null}}(A,l)=\{\textbf{0}_{n\times l}\}$.
\end{lem}
\proof
Assume $X_0$ is the unique optimal solution to problem (Mconvex).
Given a matrix $Z\in \mathcal{D}(f,X_0)\cap \mbox{null}(A,l)$,
we know that $X_0+Z$ is a feasible point of problem (Mconvex) and $f(X_0+Z)\leq f(X_0)$, which implies that $X_0+Z$ is an optimal solution to problem (Mconvex).
According to the uniqueness of optimal solution of problem (Mconvex), we have $E=0$, and thus
$\mathcal{D}(f,X_0)\cap\mbox{null}(A,l)=\{\textbf{0}_{n\times l}\}$.

Conversely, suppose $\mathcal{D}(f,X_0)\cap\mbox{null}(A,l)=\{\textbf{0}_{n\times l}\}$.
Since we know that $X_0$ is a feasible solution of problem (Mconvex),
for any matrix $Z\in \textrm{null}(A,l)\backslash\left\{\textbf{0}_{n\times n}\right\}$,
$X_0+Z$ is also feasible.
If $f(X_0+Z)\leq f(X_0)$, then we have $Z\in\mathcal{D}(f,X_0)
\cap \mbox{null}(A,l)\backslash\left\{\textbf{0}_{n\times l}\right\}=\emptyset$,
which is impossible.
Therefore
$$f(X_0+Z)>f(X_0)\mbox{ for all }
Z\in \mbox{null}(A,l)\backslash\left\{\textbf{0}_{n\times l}\right\},$$
which means that $X_0$ is the unique optimal solution to problem (Mconvex).
\qed

Since linear subspace is also a cone, Lemma \ref{ground truth} connects the optimal conditions to the relation that the intersection between the descent cone at $X_0$ and matrix null space is singleton ({\em i.e.}, problem (Mconvex) succeeds).

For a random sensing matrix $A$, the probability of success for problem (Mconvex) can be related to the ``sizes" of two cones in Lemma \ref{ground truth}.
Unfortunately, since a cone may be not linear, there's no a standard definition to describe the size of a cone.
Amelunxen {\em et al.} \cite{ALMT14} give a way to measure the size of a cone, as described in the following.

\begin{defn}
(Statistical Dimension \cite{ALMT14})\\
The statistical dimension (S.D.) $\delta(\mathcal{C})$ of a closed convex cone $\mathcal{C}\subset\R^n$ is defined as:
$$\delta(\mathcal{C}):=\mathbb{E}\left[\left\|\prod(g,\mathcal{C})\right\|_2^2\right],$$
where $g\in\R^n$ is a standard normal vector and $\d\prod(\cdot,\mathcal{C})$, denoting the Euclidean projection onto $\mathcal{C}$, is defined as: $\displaystyle \prod(x,\mathcal{C}):=\arg\min\{\|x-y\|_2:y\in \mathcal{C}\}.$
\end{defn}

According to the definition of S.D. of a cone, Amelunxen \emph{et al.} \cite{ALMT14} derive the probability that two cones with a random rotation are separated as follows.

\begin{thm}\label{kinematic formula}
\emph{(Approximate kinematic formula \cite{ALMT14})}\\
Fix a tolerance $\eta\in(0,1)$. Suppose that $\mathcal{C}_1,~\mathcal{C}_2\subset\R^N$ are closed convex cones, but one of them is not a subspace.
Draw an orthogonal matrix $Q\in\R^{n\times n}$ uniformly at random. Then
$$\delta(\mathcal{C}_1)+\delta(\mathcal{C}_2)\leq n-a_\eta\sqrt{n} ~ \Rightarrow ~ \mathbb{P}\{\mathcal{C}_1\cap Q\mathcal{C}_2=\{\textbf{0}\}\}\geq 1-\eta,$$
$$\delta(\mathcal{C}_1)+\delta(\mathcal{C}_2)\geq n+a_\eta\sqrt{n} ~ \Rightarrow ~ \mathbb{P}\{\mathcal{C}_1\cap Q\mathcal{C}_2=\{\textbf{0}\}\}\leq \eta.$$
The quantity $a_\eta:=8\sqrt{\log(4/\eta)}$.
\end{thm}

In order to satisfy the requirement of Theorem \ref{kinematic formula}, both $\Phi$ and $\Psi$ can be easily selected such that $A=\Phi\Psi$ is a Gaussian random matrix \cite{DLNT12}.
In compressive sensing, $\Phi$ and $\Psi$ are conventionally used to set as a Gaussian random matrix and orthonormal basis, respectively, so that $A=\Phi\Psi$ is also a Gaussian random matrix \cite{DLNT12}.
Let $\mathcal{C}_1=\mathcal{D}\left(f,X_0\right)$ and let $Q\mathcal{C}_2=\mbox{null}(A,l)$ with a random matrix $A=\Phi\Psi$ \cite{DLNT12}.
The probability of intersection given in Theorem~\ref{kinematic formula} can be reformulated as the probability of existence of unique optimal solution by Lemma~\ref{ground truth}, {\em i.e.},
\begin{eqnarray*}
\begin{array}{l}
\d\mathbb{P}(\mathcal{C}_1\cap Q\mathcal{C}_2=\{\textbf{0}\})
\d=\mathbb{P}(\mathcal{D}(f,X_0)\cap\emph{\mbox{null}}(A,l)=\{\textbf{0}_{n\times l}\})\\
\d=\mathbb{P}(\mbox{(Mconvex) succeeds}).
\end{array}
\end{eqnarray*}
Since the nullity of $A$ is $n-m$ almost surely, the dimension of $\mathcal{C}_2$ is $\delta\left(\mbox{null}(A,l)\right)=\dim\left(\mbox{null}(A,l)\right)=(n-m)l$.
Then, the probability that (Mconvex) succeeds can be estimated by Theorem \ref{MMVs kinematic formula}, which was derived in our earlier work \cite{HLHL15}.

\begin{thm}\label{MMVs kinematic formula}
\emph{(Phase transitions in MMVs recovery)}\\
Fix a tolerance $\eta\in(0,1)$.
Let $X_0\in\mathbb{R}^{n\times l}$ be a fixed matrix.
Suppose $A\in\mathbb{R}^{m\times n}$ has independent standard normal entries and $Y=AX_0$.
Then
\begin{eqnarray*}
\begin{array}{l}
\hspace*{-7pt}m\geq\frac{\delta(\mathcal{D}(f,X_0))}{l}+\frac{a_\eta\sqrt{nl}}{l}
\Rightarrow \mathbb{P}\left(\mbox{(Mconvex) succeeds}\right)\geq1-\eta;\\
\hspace*{-7pt}m\leq\frac{\delta(\mathcal{D}(f,X_0))}{l}-\frac{a_\eta\sqrt{nl}}{l}
\Rightarrow \mathbb{P}\left(\mbox{(Mconvex) succeeds}\right)\leq\eta,
\end{array}
\end{eqnarray*}
where the quantity $a_\eta:=8\sqrt{\log(4/\eta)}$.
\end{thm}

\section{Estimation of S.D. in (ML1P)}\label{sec:S.D.}
In Theorem \ref{MMVs kinematic formula}, $\delta(\mathcal{D}(f,X_0))$ plays an important role
to estimate the probability that (Mconvex) succeeds.
However, calculating the exact value of S.D. of a cone is still open.
In this section, we provide the bounds of S.D. of descent cone at the point $X_0$ associated with
convex function $\zeta_W(X)=\left\|X\right\|_{2,1}+\lambda\left\|X-W\right\|_{2,1}$ in problem (ML1P),
where function $\zeta_W$ is called $\ell_{2,1}$-norm with prior information.

\begin{thm}\label{SD bound}
\emph{(Error bound of S.D. in (ML1P))}\\
Let $\partial\zeta_W$ be subdifferential of $\zeta_W$.
Suppose $\partial\zeta_W(X)$ is nonempty and compact, and does not contain the origin.
Then, we have
$$
\inf_{\tau\geq0}F(\tau)-\xi(X)
\leq\delta\left(\mathcal{D}\left(\zeta_W,X\right)\right)
\leq\inf_{\tau\geq0}F(\tau),\footnote{The upper bound of S.D. (right inequality) follows Proposition 4.1 of \cite{ALMT14}.}
$$
where
$\xi(X)=\frac{2\left\|X\right\|_F\cdot\sup\left\{\left\|S\right\|_F:S\in\partial\zeta_W(X)\right\}}{\left<\partial\zeta_W(X),X\right>}$,
$$F(\tau):=F(\tau,X)
=\mathbb{E}\left[\mbox{dist}^2\left(G,\tau\cdot\partial\zeta_W\left(X\right)\right)\right]
\mbox{ for }\tau\geq0,$$
and
$G\in\mathbb{R}^{n\times l}$ is a Gaussian random matrix.

Moreover, for $k$-joint sparse matrix $X_0\in\mathbb{R}^{n\times l}$, we have
$$\inf_{\tau\geq0}F(\tau)-\frac{2(1+\lambda)\sqrt{n}}{(1-\lambda)\sqrt{k}}
\leq\delta\left(\mathcal{D}\left(\zeta_W,X_0\right)\right)
\leq\inf_{\tau\geq0}F(\tau).
$$
\end{thm}
\proof
For any given matrix $X$,
we have
$$\delta\left(\mathcal{D}\left(\zeta_W,X\right)\right)
=\mathbb{E}\left[\mbox{dist}^2\left(G,\mathcal{D}\left(\zeta_W,X\right)^{\circ}\right)\right],$$
where the distance function is
$\mbox{dist}\left(G,\mathcal{C}^{\circ}\right)
=\left\|\prod\left(G,\mathcal{C}\right)\right\|_F$
for a fixed cone $C$.
According to Corollary 23.7.1 in \cite{Rockafellar},
the polar cone can be rewrite as
$\mathcal{D}\left(\zeta_W,X\right)^{\circ}
=\bigcup_{\tau\geq0}\tau\cdot\partial\zeta_W\left(X\right)$, thus
\begin{eqnarray}\label{E_inf}
\begin{array}{l}
\d\mathbb{E}\left[\mbox{dist}^2\left(G,\mathcal{D}\left(\zeta_W,X\right)^{\circ}\right)\right]
=\mathbb{E}\left[\inf_{\tau\geq0}F_G(\tau)\right],
\end{array}
\end{eqnarray}
where
$$
F_U(\tau):=F_U(\tau,X)
=\mbox{dist}^2\left(U,\tau\partial\zeta_W(X)\right), \mbox{ for } \tau\geq0.
$$

For the upper bound of $\d\delta(\mathcal{D}(\zeta_W,X))$,
since
$$\mathbb{E}\left[\inf_{\tau\geq0}F_G(\tau)\right]
\leq\inf_{\tau\geq0}\mathbb{E}\left[F_G(\tau)\right]
=\inf_{\tau\geq0}F\left(\tau\right),$$
the result follows.

Next we aim to estimate the lower bound of $\d\delta(\mathcal{D}(\zeta_W,X))$.
By the fact that
$F_G(\tau)$ is convex on $\tau\geq0$
and continuous differentiable on $\tau>0$
(Lemma C.1 in \cite{ALMT14}),
we have
\begin{align}\label{F_G}
F_G(\tau)
\geq F_G(\tau_0)+F_G'(\tau_0)(\tau-\tau_0),
\end{align}
for any $\tau$ and $\tau_0$.

Let $\tau^*$ and $\tau_G^*$ be the minimizer of $F(\tau)$
and $F_G(\tau)$, respectively.
Since
$F(\tau)$ is strictly convex on $\tau\geq0$ and
differentiable on $\tau>0$ (Lemma C.2 in \cite{ALMT14})
the minimizer $\tau^*$ of $F(\tau)$ is unique, that is,
$$\tau^*=\arg\min_{\tau\geq0}F(\tau).$$
Then, Eq. (\ref{F_G}) can be written as
$$
F_G(\tau_G^*)
\geq F_G(\tau^*)+F_G'(\tau^*)(\tau_G^*-\tau^*).
$$
($F_G'(\tau^*)$ is the right derivative provided
$\tau^*=0$).
Then the expected value of $\inf_{\tau\geq0}F_G(\tau)$
in Eq. (\ref{E_inf})
corresponding to $G$ becomes
\begin{eqnarray*}
\begin{array}{l}
\d\mathbb{E}\left[\inf_{\tau\geq0}F_G(\tau)\right]\\
=\d\mathbb{E}\left[F_G(\tau_G^*)\right]\\
\geq\mathbb{E}\left[F_G(\tau^*)\right]
+\mathbb{E}\left[F_G'(\tau^*)(\tau_G^*-\tau^*)\right]\\
=F(\tau^*)+\mathbb{E}\left[(\tau_G^*-\tau^*)\cdot(F_G'(\tau^*)
-\mathbb{E}\left[F_G'(\tau^*)\right])\right]\\
\ \ \ \ +\mathbb{E}\left[\tau_G^*-\tau^*\right]
\cdot\mathbb{E}\left[F_G'(\tau^*)\right]\\
=F(\tau^*)+\mathbb{E}[(\tau_G-\mathbb{E}[\tau_G^*])\cdot(F_G'(\tau^*)-\mathbb{E}[F_G'(\tau^*)])]\\
\ \ \ \ +\mathbb{E}[\tau_G^*-\tau^*]\cdot F'(\tau^*)\\
\geq\inf_{\tau\geq0}F(\tau)-(\mbox{Var}[\tau_G^*]\cdot\mbox{Var}[F_G'(\tau^*)])^{1/2}\\
\ \ \ \ +\mathbb{E}[\tau_G^*-\tau^*]\cdot F'(\tau^*).
\end{array}
\end{eqnarray*}

We can see that $\mathbb{E}[\tau_G^*-\tau^*]\cdot F'(\tau^*)\geq0$ since
$F'(\tau^*)=0$ if $\tau^*>0$
and $F'(\tau^*)\geq0$ if $\tau^*=0$
(because $\tau^*$ minimize $F(\tau)$).
Therefore,
\begin{eqnarray}\label{conclusion1}
\delta(\mathcal{D}(\zeta_W,X))
\geq\inf_{\tau\geq0}F(\tau)-(\mbox{Var}[\tau_G^*]\cdot\mbox{Var}[F_G'(\tau^*)])^{1/2}.
\end{eqnarray}
Next, to compute the variance of $\tau_G^*$, we need to devise a consistent method for selecting a minimizer $\tau_U$ of $F_U$.
Introduce the closed convex cone $\mathcal{C}:=\mbox{cone}(\partial \zeta_W(X))$, and notice that
$$\inf_{\tau\geq 0}F_U(\tau)=\inf_{\tau\geq0}\mbox{dist}^2(U,\tau\cdot\partial \zeta_W(X))=\mbox{dist}^2(U,\mathcal{C}).$$
In other words, the minimum distance to one of the sets $\tau\partial \zeta_W(X)$ is attained at the point $\prod_{\mathcal{C}}(U):=\arg\min\{\|U-C\|_F:C\in\mathcal{C}\}$. As such, it is natural to pick a minimizer $\tau_U$ of $F_U$ according to the rule
\begin{equation}\label{tau_defi}
\tau_U:=\inf\{\tau\geq0:\prod_{\mathcal{C}}(U)\in\tau\partial \zeta_W(X)\}=\frac{\left<\prod_{\mathcal{C}}(U),X\right>}{\left<\partial \zeta_W(X), X\right>}.
\end{equation}
In light of Eq. (\ref{tau_defi}), we have
\begin{eqnarray*}
\begin{array}{l}
|\tau_U-\tau_V|=\frac{1}{\left<\partial \zeta_W(X), X\right>}|\left<\prod_{\mathcal{C}}(U)-\prod_{\mathcal{C}}(V),X\right>|\\
\leq\frac{\|X\|_F}{\left<\partial \zeta_W(X), X\right>}\cdot\|\prod_{\mathcal{C}}(U)-\prod_{\mathcal{C}}(V)\|_F\\
\leq\frac{\|X\|_F}{\left<\partial \zeta_W(X), X\right>}\cdot\|U-V\|_F.
\end{array}
\end{eqnarray*}
We have used the fact (B.3) in \cite{ALMT14} that the projection onto a closed convex set is nonexpansive.
By the relation between Var$(\tau_G^*)$ and Lipschitz constant
$\frac{\|X\|_F}{\left<\partial \zeta_W(X), X\right>}$, we have
\begin{equation}\label{conclusion2}
(\mbox{Var}(\tau_G^*))^{1/2}\leq\frac{\|X\|_F}{\left<\partial \zeta_W(X), X\right>}.
\end{equation}
By the lemma (C.1) in \cite{ALMT14},
\begin{equation}\label{conclusion3}
(\mbox{Var}(F_G'(\tau^*)))^{1/2}\leq2\sup_{S\in\partial \zeta_W(X)}\|S\|_F.
\end{equation}
Substitute $X_0$ into
Eqs. (\ref{conclusion1}), (\ref{conclusion2}), and (\ref{conclusion3}).
In Eq. (\ref{conclusion2}), $\left<\partial \zeta_W(X_0), X_0\right>$ can be reformulated by cosine function to $\|X_0\|_{2,1}+\lambda\sum^n_{i=1}\|x_0^i\|_2\cdot\cos(\angle Ox_0^iw_i)$.
It is obvious that the lower bound of $\left<\partial \zeta_W(X_0), X_0\right>$ is $1-\lambda\|X_0\|_{2,1}$.
In Eq. (\ref{conclusion3}), the right hand side $\sup\{\|S\|_F:S\in\partial \zeta_W(X_0)\}$ will be equal to $(1+\lambda)\sqrt{n}$ because the rows of $\partial\|X_0\|_{2,1}$ and $\partial\|X_0-W\|_{2,1}$ are already normalized.
We have
\begin{eqnarray*}
\begin{array}{l}
\delta(\mathcal{D}(\zeta_W,X_0))\\
\d\geq\inf_{\tau\geq0}F(\tau)-\frac{2\|X_0\|_F\sup\{\|S\|_F:S\in\partial \zeta_W(X_0)\}}{\left<\partial \zeta_W(X_0), X_0\right>}\\
\d\geq\inf_{\tau\geq0}F(\tau)-\frac{2(1+\lambda)\sqrt{n}\|X_0\|_F}{(1-\lambda)\|X_0\|_{2,1}}\\
\d\geq\inf_{\tau\geq0}F(\tau)-\frac{2(1+\lambda)\sqrt{n}}{(1-\lambda)\sqrt{k}},
\end{array}
\end{eqnarray*}
where the last inequality depends on $-\frac{\|X_0\|_F}{\|X_0\|_{2,1}}\geq-\frac{1}{\sqrt{k}}$. We complete the proof.\qed

\\
To calculate the function $F(\tau)$ in Theorem \ref{SD bound}, we first compute the subdifferential of both $\ell_{2,1}$-norm and $\zeta_W(X)$.

\begin{lem}\label{subdiff of L21}
\emph{(Subdifferential of $\ell_{2,1}$-norm \cite{Haltmeier13})}\\
For any $X, U\in\mathbb{R}^{n\times l}$, we have
$$U\in\partial\|X\|_{2,1}\Leftrightarrow u^i\in\partial\|x^i\|_2,
\ 1\leq i\leq n,$$
where
    \begin{equation*}
        u^i\in\partial\|x^i\|_2 \Leftrightarrow
        \begin{cases}
            u^i=x^i/\|x^i\|_2 & \text{if } x^i \neq 0,\\
            \|u^i\|_2\leq 1 & \text{if } x^i = 0.\\
        \end{cases}
    \end{equation*}
\end{lem}

The subgradient of $\ell_{2,1}$-norm at $X$ is calculated by row-by-row subgradient of Euclidean norm $\left\|\cdot\right\|_2$,
whereas $\partial\left\|x^i\right\|_2$ consists of the gradient whenever $x^i\neq0$, and $\partial\left\|x^i\right\|_2=\overline{\mathcal{B}}$ if $x^i=0$. That is, the computation of subgradient of $\ell_{2,1}$-norm at $X$ depends on if a row of $X$ is zero or not.

Moreover, since the subdifferential of $\zeta_W(X)$ can be calculated separately as
$\partial(\|X\|_{2,1}+\lambda\|X-W\|_{2,1})=\partial\|X\|_{2,1}+\lambda\partial\|X-W\|_{2,1}$,
we calculate the subgradient of $\zeta_W(X)$
according to the indices sets of zero and nonzero rows with respect to $X$ and $X-W$.
We separate the domain of $\zeta_W(X)$ into four cases, where
$E_1=\Lambda_X\cap\Lambda_{X-W}$, $E_2=\Lambda_X\cap\Lambda_{X-W}^c$, $E_3=\Lambda_X^c\cap\Lambda_{X-W}$, and $E_4=\Lambda_X^c\cap\Lambda_{X-W}^c$.
Then, we have the following lemma.

\begin{lem}\label{subdiff of L21+L21}
\emph{(Subdifferential of $\ell_{2,1}$-norm with prior information)}\\
For any $X, U\in\mathbb{R}^{n\times l}$, we have
\begin{eqnarray*}
\begin{array}{l}
U\in\partial\zeta_W\left(X\right)\Leftrightarrow
u^i\in\partial(\|x^i\|_2+\lambda\|x^i-w^i\|_2), 1\leq i\leq n,
\end{array}
\end{eqnarray*}
where
    \begin{eqnarray*}\label{subdifferential L21+L21}
        \begin{array}{l}
            u^i\in\partial(\|x^i\|_2+\lambda\|x^i-w^i\|_2)\Leftrightarrow \\
            \begin{cases}
                u^i=\frac{x^i}{\|x^i\|_2}+\lambda(\frac{x^i-w^i}{\|x^i-w^i\|_2}), &\text{if } i\in E_1,\\
                u^i=\frac{x^i}{\|x^i\|_2}+\lambda\beta^i, \|\beta^i\|_2\leq 1, &\text{if } i\in E_2,\\
                u^i=\alpha^i+\lambda(\frac{x^i-w^i}{\|x^i-w^i\|_2}), \|\alpha^i\|_2\leq 1, &\text{if } i\in E_3,\\
                u^i=\alpha^i+\lambda\beta^i, \|\alpha^i\|_2\leq 1, \|\beta^i\|_2\leq 1, &\text{if } i\in E_4.
            \end{cases}
        \end{array}
    \end{eqnarray*}
\end{lem}

According to Lemma \ref{subdiff of L21+L21}, Theorem \ref{SD bound} can be rewritten as follows.

\begin{thm}\label{S.D. of prior}
\emph{(Statistical dimension of descent cone of $\ell_{2,1}$-norm with prior information)}\\
With the same notations and assumptions as in Theorem \ref{SD bound}, the S.D. of the descent cone of $\zeta_W$ at the point $X_0$ satisfies the inequality
\begin{equation}\label{error bound}
\psi_p-\frac{2(1+\lambda)\sqrt{n}}{(1-\lambda)\sqrt{k}}
\leq\delta(\mathcal{D}(\zeta_W,X_0))
\leq\psi_p.
\end{equation}
The function $\psi_p$ is defined as $\psi_p(\textbf{E}):=\inf_{\tau\geq 0}\left\{R_p(\tau,\textbf{E})\right\}$,
where $\textbf{E}=(\left|E_1\right|,\left|E_2\right|,\left|E_3\right|,\left|E_4\right|)$ and $R_p=T_1+T_2+T_3+T_4$ with
\begin{equation*}
\begin{aligned}
T_1&=\left|E_1\right|(l+\tau^2+\tau^2\lambda^2)+2\tau^2\lambda \sum_{i\in E_1}\cos(\angle Ox_0^iw^i),\\
T_2&=\left|E_2\right|\int_{\tau\lambda}^\infty(t-\tau\lambda)^2\cdot\frac{\tau t^l e^{-\frac{t^2+\tau^2}{2}}}{(\tau t)^{l/2}}I_{l/2-1}(\tau t)dt,\\
T_3&=\left|E_3\right|\int_{\tau}^\infty(t-\tau)^2\cdot\frac{\tau\lambda t^l e^{-\frac{t^2+\tau^2\lambda^2}{2}}}{(\tau\lambda t)^{l/2}}I_{l/2-1}(\tau\lambda t)dt,\\
T_4&=\left|E_4\right|\frac{2^{1-L/2}}{\Gamma(l/2)}\int^{\infty}_{\tau(1+\lambda)}\!(t-\tau(1+\lambda))^2t^{l-1}e^{-t^2/2}dt,
\end{aligned}
\end{equation*}
where
$\Gamma$ is gamma function and \\
$\d I_{v}(z)=\sum^\infty_{k=0}\frac{1}{\Gamma(k+1)\Gamma(v+k+1)}\left(\frac{z}{2}\right)^{2k+v}$ is modified Bessel functions of the first kind.
\end{thm}
\proof
First we separate $F_G(\tau)$ as follow:
\begin{align}
&\d\mbox{dist}^2(G,\tau\cdot\partial\zeta_W(X_0))\nonumber\\
=&\d\sum_{i\in E_1}\left\|g^i-\tau\cdot\left(\frac{x_0^i}{\|x_0^i\|_2}+\lambda\left(\frac{x_0^i-w^i}{\|x_0^i-w^i\|_2}\right)\right)\right\|^2_2\label{proof step1.1}\\
&\d+\sum_{i\in E_2}\inf_{\beta^i\in\overline{\mathcal{B}}}\left\|g^i-\tau\cdot\left(\frac{x_0^i}{\|x_0^i\|_2}+\lambda\beta^i\right)\right\|^2_2\label{proof step1.2}\\
&\d+\sum_{i\in E_3}\inf_{\alpha^i\in\overline{\mathcal{B}}}\left\|g^i-\tau\cdot\left(\alpha^i+\lambda\cdot\frac{x_0^i-w^i}{\|x_0^i-w^i\|_2}\right)\right\|^2_2\label{proof step1.3}\\
&\d+\sum_{i\in E_4}\inf_{\alpha^i,\beta^i\in\overline{\mathcal{B}}}\left\|g^i-\tau\cdot\left(\alpha^i+\lambda\beta^i\right)\right\|^2_2.\label{proof step1.4}
\end{align}

In Eq. (\ref{proof step1.1}),
for each $i\in E_1$,
let $\gamma^i=\frac{x_0^i}{\|x_0^i\|_2}+\lambda\left(\frac{x_0^i-w^i}{\|x_0^i-w^i\|_2}\right)$.
By taking the expected value of Eq. (\ref{proof step1.1}),
together with the fact that $g_j^i\sim N(0,1)$,
we have
\begin{eqnarray*}
\begin{array}{l}
\d\mathbb{E}\left[\sum_{i\in E_1}\left\|g^i-\tau\gamma^i\right\|^2_2\right]\\
\d=\mathbb{E}\left[\sum_{i\in E_1}\sum^{l}_{j=1}\left(g^i_j-\tau \gamma^i_j\right)^2\right]\\
\d=\mathbb{E}\left[\sum_{i\in E_1}\sum^{l}_{j=1}\left((g^i_j)^2-2\tau \gamma^i_j g^i_j+\tau^2(\gamma^i_j)^2\right)\right]\\
\d=\sum_{i\in E_1}\sum^{l}_{j=1}\left(\mathbb{E}\left[(g^i_j)^2\right]-2\tau \gamma^i_j\mathbb{E}\left[g^i_j\right]+\tau^2(\gamma^i_j)^2\right)\\
\d=\sum_{i\in E_1}\sum^{l}_{j=1}\left(1+\tau^2(\gamma^i_j)^2\right)\\
\d=\left|E_1\right|(l+\tau^2+\tau^2\lambda^2)+2\tau^2\lambda \sum_{i\in E_1}\cos(\angle Ox_0^iw^i)\\
\d=T_1.
\end{array}
\end{eqnarray*}

In Eq. (\ref{proof step1.2}),
for each $i\in E_2$,
let $\overline{\gamma}^i=g^i-\tau\frac{x_0^i}{\|x_0^i\|_2}$,
the minimization problem can be
written as
\begin{eqnarray}\label{minimization problem of proof step1.2}
\inf_{\beta^i\in\overline{\mathcal{B}}}\left\|\overline{\gamma}^i-\tau\lambda\beta^i\right\|^2_2.
\end{eqnarray}
We can see that the optimal value is 0 provided
$\left\|\overline{\gamma}^i\right\|_2\leq\tau\lambda$.
In the case $\left\|\overline{\gamma}^i\right\|_2>\tau\lambda$,
the optimal solution is
$\beta^i=\frac{\overline{\gamma}^i}{\left\|\overline{\gamma}^i\right\|_2},$
with optimal value
$\left\|\overline{\gamma}^i\right\|_2-\tau\lambda$.
That is, the optimal value of Eq. (\ref{minimization problem of proof step1.2}) is
\begin{eqnarray*}
\begin{array}{l}
\d\inf_{\beta^i\in\overline{\mathcal{B}}}\left\|\overline{\gamma}^i-\tau\lambda\beta^i\right\|^2_2
=\left\{\begin{array}{ll}0&\mbox{if }\left\|\overline{\gamma}^i\right\|_2\leq\tau\lambda\\\left\|\overline{\gamma}^i-\tau\lambda\frac{\overline{\gamma}^i}{\left\|\overline{\gamma}^i\right\|_2}\right\|_2^2&\mbox{if }\left\|\overline{\gamma}^i\right\|_2>\tau\lambda,\end{array}\right.
\end{array}
\end{eqnarray*}
and hence Eq. (\ref{proof step1.2}) becomes
\begin{eqnarray}\label{proof step2.2}
\sum_{i\in E_2}\inf_{\|\beta^i\|_2\leq1}\left\|\overline{\gamma}^i-\tau\lambda\beta^i\right\|^2_2
=\sum_{i\in E_2}\left(\left\|\overline{\gamma}^i\right\|_2-\tau\lambda\right)^2_+.
\end{eqnarray}

Similarly to Eq. (\ref{proof step1.3}) and Eq. (\ref{proof step1.4}).
In Eq. (\ref{proof step1.3}),
for each $i\in E_3$,
let $\hat{\gamma}^i=g^i-\tau\lambda\cdot\frac{x_0^i-w^i}{\|x_0^i-w^i\|_2}$,
the minimization problem can be
written as
\begin{eqnarray*}
\inf_{\alpha^i\in\overline{\mathcal{B}}}\left\|\hat{\gamma}^i-\tau\alpha^i\right\|^2_2,
\end{eqnarray*}
which the optimal value is
\begin{eqnarray*}
\begin{array}{l}
\d\inf_{\alpha^i\in\overline{\mathcal{B}}}\left\|\hat{\gamma}^i-\tau\alpha^i\right\|^2_2
=\left\{\begin{array}{ll}0&\mbox{if }\left\|\hat{\gamma}^i\right\|_2\leq\tau\\\left\|\hat{\gamma}^i-\tau\frac{\hat{\gamma}^i}{\left\|\hat{\gamma}^i\right\|_2}\right\|_2^2&\mbox{if }\left\|\hat{\gamma}^i\right\|_2>\tau,\end{array}\right.
\end{array}
\end{eqnarray*}
and hence Eq. (\ref{proof step1.3}) becomes
\begin{eqnarray}\label{proof step2.3}
\sum_{i\in E_3}\inf_{\alpha^i\in\overline{\mathcal{B}}}\left\|\hat{\gamma}^i-\tau\alpha^i\right\|^2_2
=\sum_{i\in E_3}\left(\left\|\hat{\gamma}^i\right\|_2-\tau\right)^2_+.
\end{eqnarray}

In Eq. (\ref{proof step1.4}),
for each $i\in E_4$,
the optimal value of the minimization problem is
\begin{eqnarray*}
\begin{array}{l}
\d\inf_{\alpha^i,\beta^i\in\overline{\mathcal{B}}}\left\|g^i-\tau\cdot\left(\alpha^i+\lambda\beta^i\right)\right\|^2_2\\
=\left\{\begin{array}{ll}0&\mbox{if }\left\|g^i\right\|_2\leq\tau(1+\lambda)\\
\left\|g^i-\tau(1+\lambda)\cdot\frac{g^i}{\left\|g^i\right\|_2}\right\|_2^2&\mbox{if }\left\|g^i\right\|_2>\tau(1+\lambda),\end{array}\right.
\end{array}
\end{eqnarray*}
and hence Eq. (\ref{proof step1.4}) becomes
\begin{eqnarray}\label{proof step2.4}
\begin{array}{l}
\d\sum_{i\in E_4}\inf_{\alpha^i,\beta^i\in\overline{\mathcal{B}}}\left\|g^i-\tau\cdot\left(\alpha^i+\lambda\beta^i\right)\right\|^2_2\\
\d=\sum_{i\in E_4}\left(\left\|g^i\right\|_2-\tau(1+\lambda)\right)^2_+.
\end{array}
\end{eqnarray}

Next, we discuss the expected value of
Eq. (\ref{proof step2.2}) $\!\sim\!$ (\ref{proof step2.4}).
For Eq. (\ref{proof step2.2}),
let $S_{2,i}=\left\|\overline{\gamma}^i\right\|_2$,
for all $i\in E_2$.
Since $g^i_j\sim N(0,1)$,
$S_{2,i}$ follows the noncentral chi distribution with the same degrees of freedom $l$ and the same mean $\tau$ for all $i\in E_2$,
which implies that all $S_{2,i}$ have the same probability density function
$$\d\rho(S_{2,i}=s;l,\tau)=\frac{s^l\tau\cdot e^{-(s^2+\tau^2)/2}}{(\tau s)^{l/2}}\cdot I_{l/2-1}\left(\tau s\right).$$
By taking the expected value, we have
\begin{eqnarray*}
\begin{array}{l}
\d\sum_{i\in E_2}\mathbb{E}\left[\left(S_{2,i}-\tau\lambda\right)_+^2\right]\\
\d=\sum_{i\in E_2}\int_{\tau\lambda}^\infty(t-\tau\lambda)^2
\cdot\rho\left(t;l,\tau\right)dt\\
\d=\left|E_2\right|\int_{\tau\lambda}^\infty(t-\tau\lambda)^2
\cdot\rho\left(t;l,\tau\right)dt\\
\d=T_2.
\end{array}
\end{eqnarray*}

Similarly,
$S_{3,i}=\left\|\hat{\gamma}^i\right\|_2$
follow the noncentral chi distribution with the same degrees of freedom $l$, the same mean $\tau\lambda$, and
the same probability density function $\rho\left(S_{3,i}=s;l,\tau\lambda\right)$ for all $i\in E_3$.

Then, by taking the expected value, Eq. (\ref{proof step2.3})
becomes
\begin{eqnarray*}
\begin{array}{l}
\d\sum_{i\in E_3}\mathbb{E}\left[\left(S_{3,i}-\tau\right)_+^2\right]\\
\d=\sum_{i\in E_3}\int_{\tau}^\infty(t-\tau)^2
\cdot\rho\left(t;l,\tau\lambda\right)dt\\
\d=\left|E_3\right|\int_{\tau}^\infty(t-\tau)^2
\cdot\rho\left(t;l,\tau\lambda\right)dt\\
\d=T_3.
\end{array}
\end{eqnarray*}

For Eq. (\ref{proof step1.4}),
$S_{4,i}=\left\|g^i\right\|_2$
follow the chi distribution with the same degrees of freedom $l$,
and the same probability density function
$$\tilde\rho(S_{4,i}=s;l)=\frac{2^{1-\frac{l}{2}} s^{l-1} e^{-\frac{s^2}{2}}}{\Gamma\left(\frac{l}{2}\right)},$$
for all $i\in E_4$.
Then, Eq. (\ref{proof step2.4}) can be reformulated as:
\begin{eqnarray*}
\begin{array}{l}
\d\sum_{i\in E_4}\mathbb{E}\left[\left(\|g^i\|_2-\tau(1+\lambda)\right)_+^2\right]\\
\d=\sum_{i\in E_4}\int^{\infty}_{\tau(1+\lambda)}\!(t-\tau(1+\lambda))^2\cdot\tilde{\rho}(t;l)dt\\
\d=\left|E_4\right|\int^{\infty}_{\tau(1+\lambda)}\!(t-\tau(1+\lambda))^2\cdot\tilde{\rho}(t;l)dt\\
\d=T_4.
\end{array}
\end{eqnarray*}
Therefore,
$$
\mathbb{E}[\textrm{dist}^2(G,\tau\cdot\partial\zeta_W(X_0))]\\
=R_p(\tau,\textbf{E}),
$$
and we complete the proof.
\qed

Following Theorem \ref{S.D. of prior}, since $R_p$ is strictly convex, the infimum value can be computed by
finding the root of derivative of $R_p$.
Moreover, if we divide the inequality in Eq. (\ref{error bound}) by $n$, we can see that the error term
$\frac{2\left(1+\lambda\right)}{\left(1-\lambda\right)\sqrt{nk}}$
is inversely proportional to $n$.
That is, the error term is negligible as $n$ is large enough.
We verify Theorem \ref{S.D. of prior} in the next section.

\section{Verification}\label{sec:Verifications}
In this section, we verify our theoretical analysis about phase transition in compressive sensing via $\ell_{2,1}$-$\ell_{2,1}$ minimization, which were conducted
using the CVX package \cite{CVX}.
Based on Theorem \ref{S.D. of prior}, it's clear to see that S.D. is highly related to $\psi_p$, which is dominated by \textbf{\em E} and $\sum_{i\in E_1}\cos(\angle Ox^iw^i)$ named cosine term.
Hence, our simulations are divided into three categories:
(1) Examine how prior information, controlled by $\left|E_2\right|$, improve the performance,
(2) Verify how prior information with correct supports but imprecise values, controlled by $\left|E_1\right|$ and cosine term, affect the performance, and
(3) Examine how prior information with wrong supports, controlled by $\left|E_3\right|$, affect the performance.
All the parameters in the three simulations follow the setting described in the next subsection.

\subsection{Parameter Setting}
For parameter setting,
the signal dimension was fixed at $n=100$ and sparsity was set to $k=16$.
The number of measurement vectors was $l$.
Since there are no changes with performance when the length of a measurement vector $m$ is larger than  $\frac{n}{2}$ in all simulations, $m$ was set to range from $1$ to $\frac{n}{2}$ to focus on the phase transition of performance.
In our simulations, we construct a signal matrix $X_0\in\R^{n\times l}$ with $k$ nonzero rows and generate prior information $W$ with $k_W$ nonzero rows to satisfy $w^i=x^i$, $\forall i\in\Lambda_W\subset\Lambda_X$.

\subsection{Prior Information Controlled by $\left|E_2\right|$}
In the first simulation, $k_W$ is $4$ or $8$ and $l$ is $2$ or $5$.
The following procedure (Step 1 $\sim$ 3) was repeated $100$ times for each set of parameters, composed of $l$ and $k_W$.

\begin{description}
  \item[Step 1] Draw a standard normal matrix $A\in\R^{m\times n}$ and generate
$Y=AX_0$.
  \item[Step 2] Solve problem (ML1P) by CVX to obtain an optimal
\hspace*{+6pt}solution $X^*$.
  \item[Step 3] Declare success if $\|X^*-X_0\|_F\leq10^{-5}$.
\end{description}

\begin{figure}[h]
\begin{center}
  \includegraphics[width=0.5\textwidth, keepaspectratio]{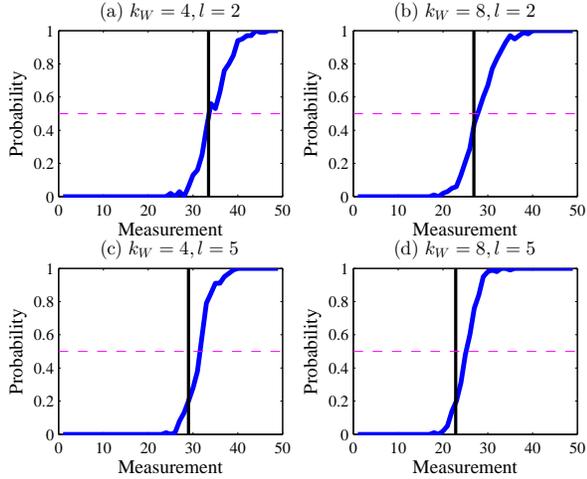}
  \caption{The empirical probability that problem (ML1P)
recovers a sparse signal matrix with the help of prior information $W$: (a) $k_W=4$ and $l=2$; (b) $k_W=8$ and $l=4$; (c) $k_W=4$ and $l=5$; (d) $k_W=8$ and $l=5$, given random linear measurements $Y=AX_0$.}
\label{exp fig1}
  \end{center}
\end{figure}

As described in Theorem \ref{S.D. of prior}, $\delta(\mathcal{D}(\zeta_W,X_0))$ depends on $n$, $\textbf{E}$, $l$, and $\lambda$.
By the definition in Theorem \ref{S.D. of prior}, $|E_1|=12$ and $|E_2|=4$ in Fig. \ref{exp fig1}(a) and (c); $|E_1|=|E_2|=8$ in Fig. \ref{exp fig1}(b) and (d). No matter $l$ equal to $2$ or $5$.
In Fig.~\!\ref{exp fig1}, the theoretical curve (in black), indicating  $\frac{\delta\left(\mathcal{D}\left(\zeta_W,X_0\right)\right)}{l}$ derived in Theorem \ref{MMVs kinematic formula}, is located at the vague region (of separating success and failure) of practical recovery results (in blue).
We can observe that the theoretical results (in black) and the practical results (in blue) in Fig.~\!\ref{exp fig1}(b) are more close to the origin than those in Fig.~\!\ref{exp fig1}(a) because
the $|E_2|$ in (b) is greater than the $|E_2|$ in (a), in other words, more correct supports ({\em i.e.}, larger $k_W$) are available.
Also we can observe that the practical result (in blue) in Fig.~\!\ref{exp fig1}(b) is better than Fig.~\!\ref{exp fig1}(a) as the former is more close to the origin.
Similar results can also be observed in Figs.~\!\ref{exp fig1}(c) and (d) when $l$ becomes larger.
In addition, they show that both the theoretical and practical results will be more close to the origin than those in Figs.~\!\ref{exp fig1}(a) and (b) due to a larger $l$ is used.
Such phenomena are reasonable because more prior information will be helpful in recovery of sparse signals.

\subsection{Prior Information with Correct Supports but Imprecise Values}
We discuss how much influence of cosine term on S.D. and performance.
This is equivalent to exploring the similarity between $X_0$ and $W$.
The parameters were $l=5$ and $k_W=8$.
We construct a matrix $X_0\in\R^{100\times 5}$ with $k=16$ nonzero rows and generate prior information $W$ with $k_W=8$ nonzero rows, where $\Lambda_W\subset\Lambda_X$ is chosen.
We repeat the procedure (Step 1 $\sim$ 3) $100$ times for four types of prior information, described as follows.
\begin{description}
  \item[Type 1.] $w^i\sim N(\textbf{0},I_{5\times 5}), ~~\forall i\in\Lambda_W$.
  \item[Type 2.] $w^i=\mbox{sign}(x^i), ~~\forall i\in\Lambda_W$.
  \item[Type 3.] $w^i=(\mu+3\sigma)\cdot \mbox{sign}(x^i), ~~\forall i\in\Lambda_W$, where $\mu$ and\\
\hspace*{+6pt}$\sigma$ are mean and standard deviation of $x^i$, respectively.
  \item[Type 4.] $w^i=x^i,~~\forall i\in\Lambda_W\subset\Lambda_X$.
\end{description}

\begin{figure}[h]
\begin{center}
  \includegraphics[width=0.5\textwidth, keepaspectratio]{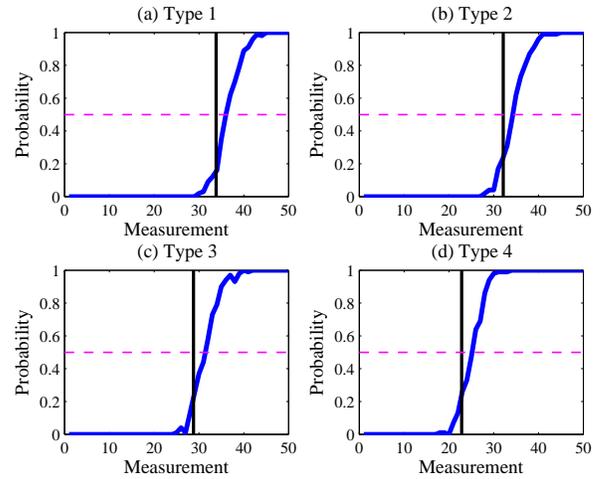}
  \caption{The empirical probability that problem (ML1P) identifies a sparse matrix with $l$ measurement vectors under prior information $W$: (a) Type 1; (b) Type 2; (c) Type 3; (d) Type 4.}
\label{exp fig2}
  \end{center}
\end{figure}

Fig. \ref{exp fig2} shows the results for four types of prior information under $n=100$ and $k_W=8$.
The results are shown in Fig. \ref{exp fig2} and are summarized as follows:
(1) As shown in Fig. \ref{exp fig2} (a), Type 1 makes the cosine term $\cos(\angle Ox^iw^i)$ unpredictable but is expected to be the highest one among the four types and cause the worst performance.
(2) In Fig. \ref{exp fig2} (b), $W$ only has correct signs, so it cannot ensure if $\cos(\angle Ox^iw^i)$ is greater than or less than $0$.
However, correct direction still improves the performance.
(3) In Fig. \ref{exp fig2} (c), $W$ has correct signs with the original signal and satisfies $|x^i_j|<|w^i_j|$ for $i\in\Lambda_W$ and $1\leq j\leq l$ with probability as high as $99\%$.
These make the cosine term less than $0$ and lead to better performance.
(4) Since Type 4 carries the best prior information, Fig. \ref{exp fig2} (d) exhibits the upper bound of performance.

\subsection{Prior Information with Wrong Supports}
\begin{figure}[!htbp]
\begin{center}
  \includegraphics[width=0.5\textwidth, keepaspectratio]{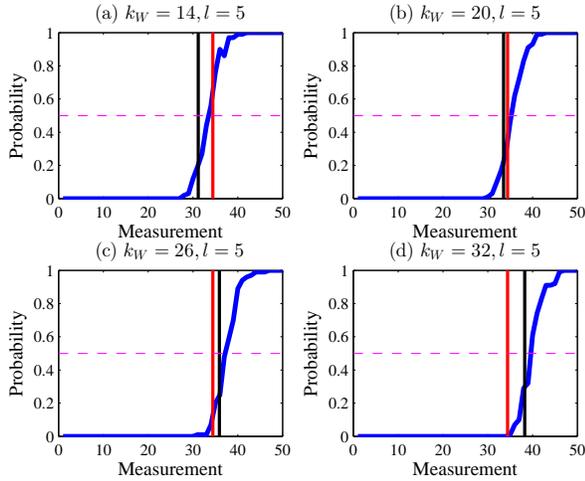}
  \caption{The empirical probability that problem (ML1P) identifies a sparse matrix with prior information $W$ and $L$ measurement vectors (a) $k_W=14$ with 6 wrong supports, (b) $k_W=20$ with 12 wrong supports, (c) $k_W=26$ with 18 wrong supports, (d) $k_W=32$ with 24 wrong supports, given random linear measurements $Y=AX_0$.}\label{exp fig3}
  \end{center}
\end{figure}

For the last simulation, we verify whether the effect of prior information with wrong supports is correctly predicted by Theorem~\ref{S.D. of prior}.
The parameters were set as $l=5$ and $k_W=8$.
Prior information with Type 3 was considered here.
Next, we choose some $i\in\Lambda_X^c$ such that $w^i\sim N(\textbf{0},I_{l\times l})$ randomly.
The procedure (Step 1 $\sim$ 3) was repeated $100$ times for each pair of parameters, $m$ and $k_W$, under four cases of different numbers of wrong supports as the prior information.
As shown in Fig. \ref{exp fig3}, they are $\left|E_3\right|=6,~\left|E_4\right|=78$ in (a), $\left|E_3\right|=12,~\left|E_4\right|=72$ in (b), $\left|E_3\right|=18,~\left|E_4\right|=66$ in (c) and $\left|E_3\right|=24,~\left|E_4\right|=60$ in (d).

To compare with the case without prior information, the results regarding $\delta(\mathcal{D}(\|\cdot\|_{2,1},X_0))$ are labeled in red line in Fig. \ref{exp fig3}.
In Fig. \ref{exp fig3} (a), although $\left|E_3\right|=6$, but it still have 8 correct supports information, overall, S.D. with such $W$ still much lower than red line.
In Fig. \ref{exp fig3} (b), $\left|E_3\right|$ increase to 12, S.D. with such $W$ become almost nothing different then red line.
In Fig. \ref{exp fig3} (c) and (d), along with the increase of $\left|E_3\right|$, the performance degrades and blue line is even greater than red line, in other words, $\ell_{2,1}$-norm minimization without prior information will gives better performance.

\section{Conclusion}\label{sec:conclusion}
In view of the fact that the phase transition analysis in joint-sparse signal recovery with prior information of compressive sensing is relatively unexplored, we have presented a new phase transition analysis based on conic geometry to figure out the effect of prior information for MMVs in this paper.
Our studies indeed provide useful insights into the critical problem of selecting prior information to guarantee improvement of signal recovery in the context of compressive sensing.

\section{Acknowledgment}
This work was supported by Ministry of Science and Technology, Taiwan (ROC), under grants MOST 104-2221-E-001-019-MY3 and 104-2221-E-001-030-MY3.

\bibliographystyle{IEEEbib}
\bibliography{refs}

\end{document}